\documentclass[%
reprint,
superscriptaddress,
 amsmath,amssymb,
 aip,
]{revtex4-1}

\usepackage{multibib}
\usepackage[colorlinks=true,%
bookmarks=false,%
linkcolor=blue,%
urlcolor=blue,%
citecolor=blue,%
breaklinks]{hyperref}

\usepackage{graphicx}
\usepackage{dsfont}
\usepackage{dcolumn}
\usepackage{bm}

\usepackage{lipsum}
\usepackage{color}
\usepackage[normalem]{ulem}
\usepackage{simplewick}
\usepackage{qcircuit}
\usepackage{braket}
\usepackage{float}
\usepackage{multirow}
\usepackage{tikz}
\usepackage[utf8]{inputenc}

\usepackage{braket}
\usepackage{subfig}

\usepackage{enumitem}

\begin{document}

\title{Extending the reach of quantum computing for materials science with machine learning potentials }

\author{Julian Schuhmacher}
\affiliation{IBM Quantum, IBM Research – Zurich, 8803 Rüschlikon, Switzerland}
\affiliation{Institute for Theoretical Physics, ETH Zürich, 8093 Zürich, Switzerland}
\author{Guglielmo Mazzola}
\email{gma@zurich.ibm.com}
\author{Francesco Tacchino}
\affiliation{IBM Quantum, IBM Research – Zurich, 8803 Rüschlikon, Switzerland}
\author{Olga Dmitriyeva}
\affiliation{IBM Quantum, Yorktown Heights, NY, US}
\author{Tai Bui}
\author{Shanshan Huang}
\affiliation{Applied Sciences, Innovation and Engineering, BP International Ltd., UK}
\author{Ivano Tavernelli}
\email{ita@zurich.ibm.com}
\affiliation{IBM Quantum, IBM Research – Zurich, 8803 Rüschlikon, Switzerland}

\date{\today}

\begin{abstract}
  Solving electronic structure problems represents a promising field of application for quantum computers.
  Currently, much effort has been spent in devising and optimizing quantum algorithms for quantum chemistry problems featuring up to hundreds of electrons.
  While quantum algorithms can in principle outperform their classical equivalents, the polynomially scaling runtime, with the number of constituents, can still prevent quantum simulations of large scale systems.
  We propose a strategy to extend the scope of quantum computational methods to large scale simulations using a machine learning potential, trained on quantum simulation data.
  The challenge of applying machine learning potentials in today's quantum setting arises from the several sources of noise affecting the quantum computations of electronic energies and forces. 
  We investigate the trainability of a machine learning potential selecting various sources of noise: statistical, optimization and hardware noise. 
  Finally, we construct the first machine learning potential from data computed on actual IBM Quantum processors for a hydrogen molecule.
  This already would allow us to perform arbitrarily long and stable molecular dynamics simulations, outperforming all current quantum approaches to molecular dynamics and structure optimization.

\end{abstract}

\maketitle

\section{Introduction}

In view of the exponential speed-up that can be achieved in solving electronic structure problems
quantum computers have the potential to revolutionize the field of quantum chemistry~\cite{Feynman1982,aspuru2005simulated,Moll_2018,Cao_2019}. 
Substantial experimental and theoretical advances have been made in the last years, both concerning the realization of quantum computing platforms~\cite{Krantz_2019,bruzewicz2019trapped,Chatterjee2021,Saffman_2010},
and the development of new generations of quantum algorithms~\cite{montanaro2016quantum,Bauer_2020,cerezo2021variational,bharti2021noisy}.

In a nutshell, by means of quantum computers one can in principle accurately solve the Schr\"odinger equation with an algorithm that scales polynomially in both, memory and runtime, while exact classical configuration interaction methods scale exponentially with the system size.
All the various quantum algorithms proposed to solve this problem, from
eigenstate projection methods using quantum phase estimation~\cite{abrams1999quantum,aspuru2005simulated}, to  variational approaches~\cite{Peruzzo2014,Kandala2017}, feature a $\mathcal{O}(N^4)$ scaling where $N$ is essentially the system size (or more specifically the number of basis functions used to represent the system).

The origin of this scaling is the number, $\mathcal{O}(N^4)$, of terms present in an electronic Hamiltonian in second quantization~\cite{Cao_2019}.
This means that the complexity of one Trotter step in the Hamiltonian evolution primitive also scales with $\mathcal{O}(N^4)$. 
Concerning the variational approach, the number of measurements needed to compute the energy necessarily scales with the number of terms in the Hamiltonian~\cite{wecker2015progress}. 
Moreover, the most promising variational circuit to represent chemical systems, the unitary coupled cluster (UCC) ansatz, also feature a complexity of $\mathcal{O}(N^4)$ gates, for the same reason.

Much of the effort spent so far has been focused on the solution of medium-sized chemical systems, while virtually no attempts have been made to propose a feasible strategy for simulations of extended systems, central in materials science.

Crucially, the asymptotic exponential speed-up that a quantum computation can offer, may not be sufficient to simulate bulk systems.
While end-to-end resource estimates confirm that quantum computers equipped with optimized quantum algorithms can perform chemistry calculations that are completely unfeasible for current classical solvers~\cite{Reiher_2017,von2020quantum}, the $\mathcal{O}(N^4)$ scaling can still represent a practical barrier to achieve quantum speed-up in first-principle simulations of extended systems.

For instance, approximate classical electronic solvers like Density Functional Theory (DFT)~\cite{Burke_2012} or quantum Monte Carlo~\cite{RevModPhys.73.33,nakano2020turborvb}, which feature a $\mathcal{O}(N^2)-\mathcal{O}(N^3)$ scaling, stand at the edge of what we can really consider large scale simulations.
Indeed, first-principle molecular dynamics (MD) simulations, powered by DFT, can routinely tackle systems featuring an order of $10^3$ electrons on picosecond (ps) time scales.
These sizes and timescales (not to speak of accuracy) are often not sufficient to realize a converged setup to study many physical systems, ranging from chemical reactions in solutions, nucleation processes, and phase transitions.

On top of that, the quantum gate frequency is typically orders of magnitude slower than on a classical CPU~\cite{gidney2019efficient,Reiher_2017}.
This means that a large prefactor should also be taken into account when considering the runtime of a quantum algorithm.
All in all, it seems unlikely that an exact quantum powered electronic structure method, featuring a runtime of $\mathcal{O}(N^4)$, will achieve large scale simulations of bulk systems, whereas a classical approximate solver like DFT showing a better scaling, $\mathcal{O}(N^3)$, as well as a much better prefactor, struggles.

Furthermore, the ``time'' dimension of the problem will likely constitute a major bottleneck for first-principle simulations powered by quantum electronic structure solvers, as much as it is for present classical ones.
An MD simulation implies that a sequence of electronic structure calculations needs to be performed serially~\cite{allen_computer_1987}, each for a new generated displacement, and the total runtime of the numerical experiments also grows with $n_T$, the number of time steps. 
Given that the typical integration time step of an MD simulation for molecular systems is $0.2$ femtosecond (fs), $n_T$ should be at least of order $10^4$ for a ps long dynamics.
However, that can still be too short to produce predictive results for many condensed matter problems of interest.
For instance, phase transitions and complex chemical reactions may take place on a timescale of nanoseconds or more.

Recently, machine learning (ML) solutions have been put forward to overcome such size and timescale barriers in first-principle simulations driven by DFT electronic solvers\cite{rupp2015machine,QUA:QUA24927,perspectiveMLforcefield,QUA:QUA24890,noe2020machine}.
The last two years witnessed an exceptional increase in quantity and quality of ML-powered numerical experiments, such that this approach is on the path to become a standard in materials science~\cite{bartok,Cheng_2020,Gartner_2020,Deringer_2021}.

In this manuscript, we propose that first-principle quantum computing simulations of materials should follow the same approach. 
Namely, data coming from quantum hardware should be used to harvest high-quality electronic structure datasets to generate a machine learning potential (MLP), rather than to drive an MD trajectory directly.

However, the combination of the two approaches is less straightforward compared to the case where the dataset is generated through DFT calculations, as present quantum computation is subject to several forms of noise, which will impact the quality of the dataset.
In this work, we focus on three types of quantum noise: \emph{(i)} the statistical noise, closely connected to the problem of measurements in quantum mechanics; \emph{(ii)} the variational optimization noise, which is inherent to variational approaches; and \emph{(iii)} the hardware noise, which originates from device errors.

The manuscript is structured as follows. 
Section~\ref{s:mlp} reviews the general idea of MLPs. 
Section~\ref{s:qc} introduces the general concepts for quantum electronic structure calculations.
In section~\ref{s:errors} we discuss the effect of the different noise sources on quantum computations.
Section~\ref{s:nnff} provides a brief description of the applied Neural-Network force-field approach.
We present and discuss the results in section~\ref{s:results} and conclude in section~\ref{s:conclusion}.

\section{Machine learning potentials}
\label{s:mlp}

Machine learning approaches are the cornerstone in many technological applications, ranging from image/speech recognition, search and recommendation engines, to data detection and filtering~\cite{Goodfellow-et-al-2016}.

While in the past ML methods, due to their power of compressing high-dimensional data into low-dimensional representations\cite{hinton_reducing_2006,lecun_deep_2015}, have been mostly applied to data science, we have recently witnessed an increased interest for applications in the physical sciences, and particularly in quantum mechanics\cite{carleo2019machine,noe2020machine}.
For instance, several ML methods have been put forward to solve the many-body Schr\"odinger equation (in a reinforcement learning fashion)\cite{carleo2017solving,choo2020fermionic,pfau2020ab,hermann2020deep}, to learn quantum states from measurements (unsupervised learning)\cite{torlai2017many,rem2019identifying,rodriguez2019identifying,Torlai_2020}, or to learn materials or chemical properties from datasets (supervised learning)\cite{bartok,noe2020machine}.
The general idea in these approaches is to search large databases for non-trivial relationships between the molecular or crystal structures  (i.e. atomic positions and nuclear charges $\{{\bf R,Z}\}$) and several properties of interest. 
These include, for instance, semiconductor's band-gaps and dielectric constants\cite{pilania}, atomization energies of organic molecules\cite{PhysRevLett.108.058301,doi:10.1021/acs.jctc.5b00099}, energy formations of crystals\cite{PhysRevLett.117.135502}, or the thermodynamic stability of solids and liquids\cite{doi:10.1021/acs.chemmater.7b00156,cheng2019ab}.
To do so, one first performs the training of the ML algorithm on a finite subset of known solutions ($ \{{\bf R,Z} \} \rightarrow p$), and then predicts the properties $p$ of interest for new, unseen structures, which differ in composition and geometry.
 
The construction of MLPs, first pioneered by Blank et al.~\cite{Blank_1995}, and later by Behler $\&$ Parrinello\cite{Behler_2007,Behler_2015} falls within this class.

In essence, the approach works as follows: \emph{(i)} one generates a training dataset of $M$ configurations $\{{\bf R,Z}\}$. 
Each sample contains atomic positions (the number of atoms in the sample can be limited, provided that interactions are local in space), charges and the energy $E$ (or forces {\bf f}) calculated with an \emph{ab initio} electronic structure method, such as DFT.
\emph{(ii)} The learning process consists of generalizing the mapping $\{{ \bf R,Z}\} \rightarrow E$ to out-of-sample configurations, bypassing the need of solving the electronic structure problem at each MD iteration, thus achieving a considerable speed-up in simulations.

As a result, first-principle modelling can now reach size and time scales, which were previously possible only with computationally cheap, but approximated, empirical force-fields.
The technique has already been applied to several long-standing problems in materials modelling, such as the phase diagram of liquid and solid water\cite{cheng2019ab,Gartner_2020}, silicon\cite{Bonati_2018,Deringer_2021}, and dense hydrogen\cite{Cheng_2020}, to name a few.

In this work, we adopt the MLP approach based on a neural-network architecture\cite{Behler_2007}, as implemented in the software package \texttt{n2p2} (version: 2.1.1)~\cite{n2p2}, in combination with a training based on a quantum computing evaluation of the electronic structure. 
Details of the set-up will be outlined in Sect.~\ref{s:nnff}.
Before moving already to this rather technical part, let us discuss first the quantum computing aspect of the work.

\section{Quantum computing for electronic structure problems}
\label{s:qc}
The starting point of most electronic structure problems in chemistry or materials science is the electronic Hamiltonian written in the second quantized representation~\cite{Barkoutsos2018,Moll_2018,Cao_2019},
\begin{align}
\hat{H} (\bm{R}) &=\sum_{rs} h_{rs}(\bm{R})\, \hat{a}^{\dagger}_{r} \hat{a}_{s} \label{eq:H}  \\ &+
\frac{1}{2}\sum_{pqrs} g_{pqrs}(\bm{R}) \,   \hat{a}^{\dagger}_{p} \hat{a}^{\dagger}_{q} \hat{a}_{r} \hat{a}_{s} + E_{nn}(\bm{R}), \notag
\end{align}
with $h_{rs}(\bm{R})$ denoting the one-electron integrals and $g_{pqrs}(\bm{R})$ the two-electron integrals, respectively.
The vector of nuclear coordinates $\bm{R}= ( \bm{R}_1$,$\bm{R}_2$,...,$\bm{R}_{N_I} ) \in \mathbb{R}^{3N_I}$ of $N_I$ nuclei parameterizes the electronic Hamiltonian. 
The operators $\hat{a}_{r}^{\dagger}$~($\hat{a}_{r}$) represent the fermionic creation~(annihilation) operators for electrons in  $N$ molecular spin-orbitals (MOs).
The term $E_{nn}(\bm{R})$ represents the classical nuclear repulsion energy.

The implementation of Eq.~\eqref{eq:H} requires the translation of 
each fermionic operator into a qubit operator, that can be interpreted by a quantum computer.
This can be achieved by several fermion-to-qubit mappings such as the Jordan-Wigner or the Bravy-Kitaev mapping (we refer to standard reviews such as Refs~\cite{Barkoutsos2018,Moll_2018,Cao_2019} for more details).
After this mapping, the Hamiltonian operator has the following form
\begin{equation}
    \label{eq:paulis_decomposition}
    \hat{H} = \sum_{k=1}^{K} c_k \hat{P}_k\, , \hspace{1cm} c_k \in \mathbb{R} \, ,
\end{equation}
where each $N$-qubit Pauli string $\hat{P}_k$ is an element of the set $\mathcal{P}_N = \lbrace \hat{p}_1 \otimes \hat{p}_2 \otimes \dots \otimes \hat{p}_N~|~\hat{p}_i \in \lbrace \hat{I}, \hat{X}, \hat{Y}, \hat{Z} \rbrace \rbrace$ (tensor products of $N$ single qubit Pauli operators).
As discussed above, the total number of Pauli terms scales as $\mathcal{O}(N^4)$.

There exists several methods to solve for the ground state of Eq.~\eqref{eq:H} using a quantum computer.
The method of choice, when fault-tolerant quantum computers will be available, is to perform quantum phase estimation to project onto eigenstates of the Hamiltonian~\cite{nielsen_chuang_2010, abrams1999quantum,Cao_2019}.
Another strategy is to obtain a variational approximation of the ground state using the variational quantum eigensolver (VQE)\cite{Peruzzo2014,Cao_2019}.
This heuristic method features parameterized quantum circuits, defined in terms of parametric gates.  
This generates a variational quantum state $ | \Psi (\bm{\theta}) \rangle$, often called \emph{trial state}, defined by the array of parameters $\bm{\theta}$.
The parameters are then optimized \emph{classically} to reach the minimum for the energy
\begin{equation}
\label{e:expect}
E(\bm{\theta}) = \langle \Psi (\bm{\theta})| \hat{H}(\bm{R})| \Psi (\bm{\theta})\rangle \,.
\end{equation}

The expectation value in Eq.~\eqref{e:expect} is calculated as the sum of the  expectation values $ \langle \hat{P}_k \rangle$ of the single Pauli operators, multiplied by the respective scalar coefficients $c_k$.
Each $ \langle \hat{P}_k \rangle$ value is obtained through sampling from the prepared state $ |\Psi \rangle$ using $S_k$  measurements, hence $S_k$ repetitions of the same circuit (see Ref.~\onlinecite{Kandala_2017} for details).
The statistical error associated with the evaluation of $ \langle \hat{P}_k \rangle$ decreases as $1 / \sqrt{S_k}$.

Finally, to construct a training dataset using the VQE algorithm, one just needs to create the second quantized Hamiltonian Eq.~\eqref{eq:H} for a set of generated atomic structures $\{{\bf R,Z}\}$, perform the fermion-to-qubit mapping to generate the qubit Hamiltonian Eq.~\eqref{eq:paulis_decomposition}, and finally optimize a parameterized circuit $ | \Psi (\bm{\theta(\{R,Z\})}) \rangle$ to obtain the energies of the atomic structures.

\section{Error sources and training with noisy datasets}
\label{s:errors}

The combination of VQE with ML for force-field generation would be trivial if not for the presence of error sources that are absent in classical DFT calculations.
These errors pertain only to the quantum nature of the hardware, and while they can be systematically reduced, some of them will remain finite, assuming practical set-ups.
 
In this manuscript, we consider three main noise sources that can affect the quality of the dataset generated with a quantum computer.
 
\subsection{Statistical noise in expectation values}
\label{ss:stat_err}

This type of noise is linked with the way observables are computed in the quantum setting. As discussed above for the case of the energy, the expectation value of an operator is computed as the sum of the expectation values of its Pauli terms.
The variance becomes,
\begin{equation} \label{eq:var_H}
    \textrm{Var} \left[ H \right] = \sum_{k=1}^{K} \left| c_k \right|^2 \textrm{Var} [ \hat{P}_k ] \, ,
\end{equation}
where $\textrm{Var} [ \hat{P}_k ] = \langle P_k^2\rangle- \langle P_k\rangle^2$ is the variance of the Pauli string $\hat{P}_k$.
It is easy to see that the variance is always finite, even if we consider the exact ground state of $H$. 
Since the ground state is an eigenstate of the sum, but not of each single Pauli operator $P_k$, the total variance is always positive\cite{Wecker_2015}.
The error in the estimation is therefore given by
\begin{equation}
 \epsilon_{\mathrm{stat}}=\sqrt{\sum\limits_k \vert c_k \vert^2 \mathrm{Var}(\hat{P}_k)/S_k},
\end{equation}
where $S_k$ is the number of measurements, used to estimate the $k$-th term, with $\sum_{k=1}^K S_k = M$, and $M$ the total number of measurements.
For instance, for an 8 qubit Hamiltonian operator representing the $H_2$ molecule at equilibrium bond distance in the 6-31g basis set, the number of shots $M$ required to compute the energy within chemical accuracy ($1.6$ mHa) is on the order of $10^8$~\cite{Torlai_2020}.

As of today, many strategies have been put forward to at least mitigate this issue~\cite{Torlai_2020,huang2020predicting,hadfield2020measurements,Jena2019,Yen2019b,Huggins2019,Gokhale2019,Crawford2019,zhao2019measurement,hamamura_efficient_2020,garcia2021learning}. 
To the best of our knowledge these methods can save at most three orders of magnitude in the number of shots\cite{Torlai_2020}, but cannot remove entirely the problem.

Without loss of generality, we can therefore assume that the expectation value of an operator $O$ decomposed as a sum of Pauli terms (Eq.~\eqref{eq:paulis_decomposition}) will always take the form 
\begin{equation}
\label{eq:error1}
    \langle O \rangle = \langle O \rangle_{\mathrm{true}} + \epsilon_{\mathrm{stat}}
\end{equation}
even if the exact ground state can be represented by the quantum circuit.
The operators of our interest are the energy $E$ and the set of $N_I$ atomic forces $\mathbf{F}=(\mathbf{F}_1,\mathbf{F}_2,\dots, \mathbf{F}_{N_I})$, that can also be decomposed into Pauli strings and measured alongside the energy\cite{Sokolov_2021}.
This labeling error needs to be taken into account when training an ML model.

Finally, we notice that statistical errors in the expectation values of energy and forces are present also in some classical electronic solvers like quantum Monte Carlo\cite{tirelli2021high}.

\subsection{Variational and optimization error}
\label{ss:opt_err}
The second type of error we consider is the variational error. This happens because the exact ground state generally lies outside the region of the Hilbert state that can be represented by the variational ansatz.

So far, different types of circuits have been employed in quantum computing calculations. 
They can be roughly divided into two classes. 
The first class contains \emph{chemically inspired} circuits. These circuits generally feature few variational parameters, but a fairly large depth, and therefore are still unsuited for present day's hardware. 
The most popular of them is the unitary coupled cluster (UCC) circuit\cite{Cao_2019}, with a depth growing as $\mathcal{O}(N^4)$, in the commonly used version where the excitations are truncated at the second order (UCCSD).

The second class consists of \emph{hardware efficient} ansatze\cite{Kandala2017}. 
These circuits prepare entangled states while minimizing the circuit depth. 
They usually feature many more variational parameters, therefore they offload part of the computational burden to the classical optimizer.

Indeed, even in the case when the exact ground state, or a close approximation of it, is theoretically within the representability range of the ansatz, suboptimal energy minimization can lead to poor results.
Eq. \eqref{eq:error1} is thus modified as
\begin{equation}
\label{eq:error2}
    \langle O \rangle = \langle O \rangle_{\mathrm{true}} + \epsilon_{\mathrm{stat}} + \epsilon_{\mathrm{var}}
\end{equation}
where $\epsilon_{\mathrm{var}}$ is the error coming from a non-ideal variational optimization.
In this work, we will study how this error depends on the chosen circuit and how it impacts the training of an MLP.

\subsection{Hardware noise}
\label{ss:hw_err}
In the era of noisy quantum devices, errors occur in the execution of a quantum circuit on actual quantum processors. 
As a result, data sets prepared via quantum computing methods will be affected by inaccuracies even in the ideal case of a perfect choice of the ansatz (see Sec.~\ref{ss:opt_err} above). 
It is therefore important to assess the possibility of successfully training a good MLP even in the presence of these effects. 
Incoherent errors and readout noise, which may increase fluctuations and bias in the energy evaluations and can even hinder the optimization of VQE ansatzes~\cite{wang2020noiseinduced}, are particularly important in this context. 

Errors belonging to the first class, namely incoherent noise, are primarily due to unwanted and uncontrolled interactions between qubits and their environment throughout the whole computation. 
These formally translate into finite relaxation and coherence times, named $T_1$ and $T_2$ respectively, which essentially correspond to amplitude and phase damping effects. 

Readout errors instead affect the qubit measurement process: these may be modelled as bit flip channels which stochastically produce erroneous assignments while the state of a qubit is being probed.

Finally, coherent errors may also arise in the implementation of single- and 2-qubit logic gates, primarily due to imperfect device calibration and manipulation. These typically result in systematic infidelities of the individual operations. \\

Two observations are in order. 
On the one hand, it would seem cautious to expect that standard ML techniques will not be able by themselves to compensate for hardware noise, unless specifically designed for this purpose~\cite{Kim2020ieee,bennewitz2021neural,Cincio2021mlnoise}: as a result, a minimal well posed target is to show trainability of an MLP up to an overall model error -- with respect to noiseless exact values -- matching as close as possible the characteristic inaccuracy induced by noise on the training points. 
On the other hand, one should also keep in mind that fast technological advancements, possibly in combination with error mitigation techniques~\cite{Li_error_mit_prx_2017,kandala_error_2019,endo_practical_2018,bravyi2020mitigating,suchsland2020algorithmic,Huggins2019,koczor2020exponential,mcclean_decoding_2020,obrien_prxq_2021,chen_exponential_2021}, will progressively reduce the impact of hardware noise. 
It is therefore interesting to investigate the possible improvements that ML generated potentials could enjoy in the future, showing that their quality could closely follow the increased accuracy of the available datasets.

\section{Neural-network force fields}
\label{s:nnff}
In this paper, we adopt the high-dimensional neural-network potential (HDNNP) architecture of Behler and Parinello\cite{Behler_2007} for the machine learning potential (MLP).
For the general motivation and the description of this ML model we refer to a review by J\"org Behler~\cite{Behler_2015}. 
Here, we provide a detailed discussion of some non-trivial aspects of the architecture, which are also important to reproduce our results.

We use the following procedure for the training of an MLP.
\emph{(i)} Prepare a training and a validation dataset. 
This should be done with VQE as explained in Sect.~\ref{s:qc}.
\emph{(ii)} Fix the neural-network architecture (neural-network geometry, learning parameters, symmetry functions)
\emph{(iii)} Train an MLP using the training dataset.
\emph{(iv)} Evaluate the MLP on the validation dataset. 
\emph{(v)} Repeat steps \emph{(ii-iv)} for different sets of hyperparameters.
\emph{(vi)} Choose the MLP with the lowest prediction error on the validation dataset.
  
The prediction error is measured in terms of the \textit{root-mean-square error} (RMSE), which is defined for the energy (E) as
\begin{equation}\label{eq:rmse_energy}
    \text{RMSE}(E) = \sqrt{ \frac{1}{N_{\text{s}}} \sum\limits_{i=1}^{N_{\text{s}}} \left( E^i_{\text{MLP}} - E^i_{\text{Ref}} \right)^2 }
\end{equation}
and for the forces (F) as
\begin{equation}\label{eq:rmse_forces}
    \text{RMSE}(F) = \sqrt{ \frac{1}{N_{\text{s}}} \sum\limits_{i=1}^{N_{\text{s}}} \frac{1}{3 N^i_a} \sum\limits_{j=1}^{N^i_a} \left( {\bf F}^{i}_{\text{MLP},j} - {\bf F}^{i}_{\text{Ref},j} \right)^2 } \,,
\end{equation}
where the explicit dependence of $N^i_a$ on the sample index $i$  comes from the fact that in general the dataset contains structures with different number of atoms.
Notice that the dataset labels are normalized as explained in Appendix~\ref{app:dataset_normalization}.

Another crucial ingredient for this procedure are the \emph{symmetry functions}.
These are many-body functions that capture, in a compact fashion, the structural information in the local environment of an atom.
The symmetry functions values are the real inputs for the NN, instead of the raw Cartesian coordinates of the atoms. 
The main motivation behind this choice is that translational and rotational invariance can be easily implemented\cite{Behler_2015}.

In this work, we adopt so-called $G_2$ and $G_3$ symmetry function classes. 
The first is a family of \emph{radial} symmetry functions made of two-body terms, while the second contains also three-body terms, which are needed to encode the tridimensional structure of an atomic configuration.
We provide in Appendix~\ref{app:symfunc} the explicit functional form of these functions, as well as other details needed to reproduce our settings.

It is important to notice that one would like to avoid redundancies in the symmetries function set. 
In this work, we first define a set of candidate symmetry functions, then select a small subset that still enables us to capture the structural information of a given dataset.
To this end, we adopt the automatic selection of symmetry functions as proposed by Imbalzano et al.~\cite{Imbalzano_2018}, which is detailed in Appendix~\ref{app:symfunc_select}.

However, in the case of the bulk systems of Sect.\ref{ss:statnoiseresult} which have already been studied in Refs~\cite{Morawietz_2016,Cheng_2020}, we adopt the symmetry functions already used in the respective publications.

\section{Results and Discussion}
\label{s:results}
We now present the results of the proposed HDNNP approach trained with electronic structure calculations performed with a quantum algorithm, and affected by typical noise sources compatible with near-term quantum computers. 
We proceed systematically by analyzing the impact of each different noise source on quality of the predictions for a series of model systems.
For the statistical error analysis, we start with the study of the effect of a Gaussian distributed noise model on the energies of forces evaluated for liquid and solid water. 
We then proceed to the investigation of a smaller system, namely a single water molecule, which can be implemented on today's quantum devices and for which a resource assessment is possible.
Finally, we validate our approach for the case of the H$_2$-H$_2$ cluster, where the sampling of intermolecular distances and orientations is required.
The analysis of the impact of the optimization errors on the quality of the HDNNP predictions is investigated for the same water molecule system introduced above, using different wavefunction ansaetze.
Finally, the effect of hardware noise is investigated on the simpler molecular system, namely H$_2$, for which we can efficiently perform the required sampling of the intramolecular distance both in simulations and in hardware experiments. 
Details on the simulation parameters and system setups will be introduced in order of appearance in the following sections.  

\subsection{Resilience against statistical noise}
\label{ss:statnoiseresult}
\subsubsection{A bulk system example: liquid and solid water}
The first of our assessments concerns the trainability of an MLP in the presence of the statistical noise alone (see~\ref{ss:stat_err}).
This study can be performed already in a prototypical bulk system, which is the end goal of the whole technique.
Indeed, the statistical noise in the labels of the training dataset can be easily and rigorously \emph{emulated} by adding a Gaussian distributed random variable with zero mean.
For each structure in the training dataset the reference energy and forces are modified according to
\begin{align}\label{eq:nnp_label_noise}
    E &\rightarrow E + \mathcal{N}(0, \Delta_E) \\
    F_i^\mu &\rightarrow F_i^\mu + \mathcal{N}(0, \Delta_F) \,,
\end{align}
where $E$ is the energy of the structure and $F_i^\mu$ is the force corresponding to atom $i$ and component $\mu=\{x,y,z\}$. 
$\Delta_E$ and $\Delta_F$ correspond to the variance of the statistical noise that is introduced for the energy and the forces, respectively.

In this study, we consider a bulk water system.
The dataset is taken from Ref.~\onlinecite{Morawietz_2019} and contains 7241 configurations of ice and liquid water.
The energies and forces were calculated with DFT using the RPBE functional~\cite{Hammer_1999} with D3 corrections~\cite{Grimme_2010}.
The mean energy in the dataset is -694.47 eV/atom with a standard deviation of 0.11 eV/atom. 
The standard deviation of the forces is 1.225 eV/\AA.
Here we follow the reasonable assumption that the potential energy surface obtained with a DFT model is in qualitative agreement with the exact one, and that the remaining difference does not play any role in this particular assessment concerning the learnability of an MLP from a noisy dataset.

\begin{figure*}[ht]
    \includegraphics[width = 1 \textwidth]{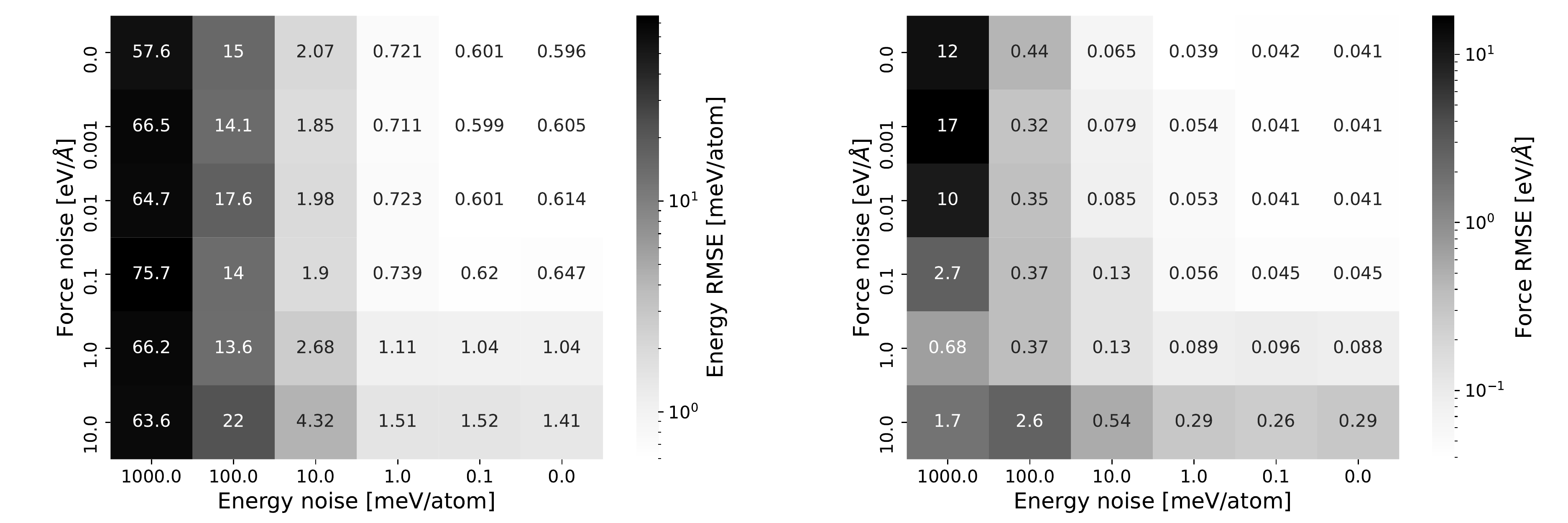}
    \caption{Bulk water system: RMSE of the MLPs trained using noisy labels. \textbf{Left} Energy RMSE as function of the noise level (see text) affecting the energy labels (x-axis) and the force labels (y-axis) in the training dataset. The standard noiseless case correspond to the top-right entry. \textbf{Right} Same assessment but targeting the MLP force RMSE.}
    \label{fig:bulkwater}
\end{figure*}

We then use the noisy training datasets to fit the MLPs and a noiseless validation dataset to assess their accuracy.
The amount of noise in the energies and the forces is varied independently. 
The values considered for the energy noise are \{1000, 100, 10, 1, 0.1\} meV/atom.
For the force noise the values \{10, 1, 0.1, 0.01, 0.001\} eV/\AA\ are used.
As a reference also the training with no noise in the energy and/or the force labels is considered.
The symmetry functions are taken from Ref.~\onlinecite{n2p2}. 
The resulting prediction RMSE values of the trained MLPs are shown in Fig.~\ref{fig:bulkwater}.

First of all, we observe that the prediction accuracy for the training on the dataset without noise (cell in the top right corner in the plots of Fig.~\ref{fig:bulkwater}) is consistent with Ref.~\onlinecite{Singraber_2019_training} (0.7 meV/atom for the energy and 0.036 eV/\AA\ for the forces). 
Most importantly, we notice that there exists a limiting value of $\Delta_E$ and $\Delta_F$ below which the prediction accuracy is as low as in the noiseless case.
The important consequence is that one is not forced to reduce the statistical error bars in the dataset to \emph{zero}, enabling in principle a practical implementation of the method.

\subsubsection{A single water molecule}
The goal of this section is to introduce a smaller system, a single water molecule, for which a quantum resource assessment is feasible.
We will translate the error threshold $\Delta_E$ into a quantum measurement resource estimate.
The single molecule configurations are extracted from the bulk water dataset used in the previous section. 

\begin{figure*}[ht]
    \includegraphics[width = 1 \textwidth]{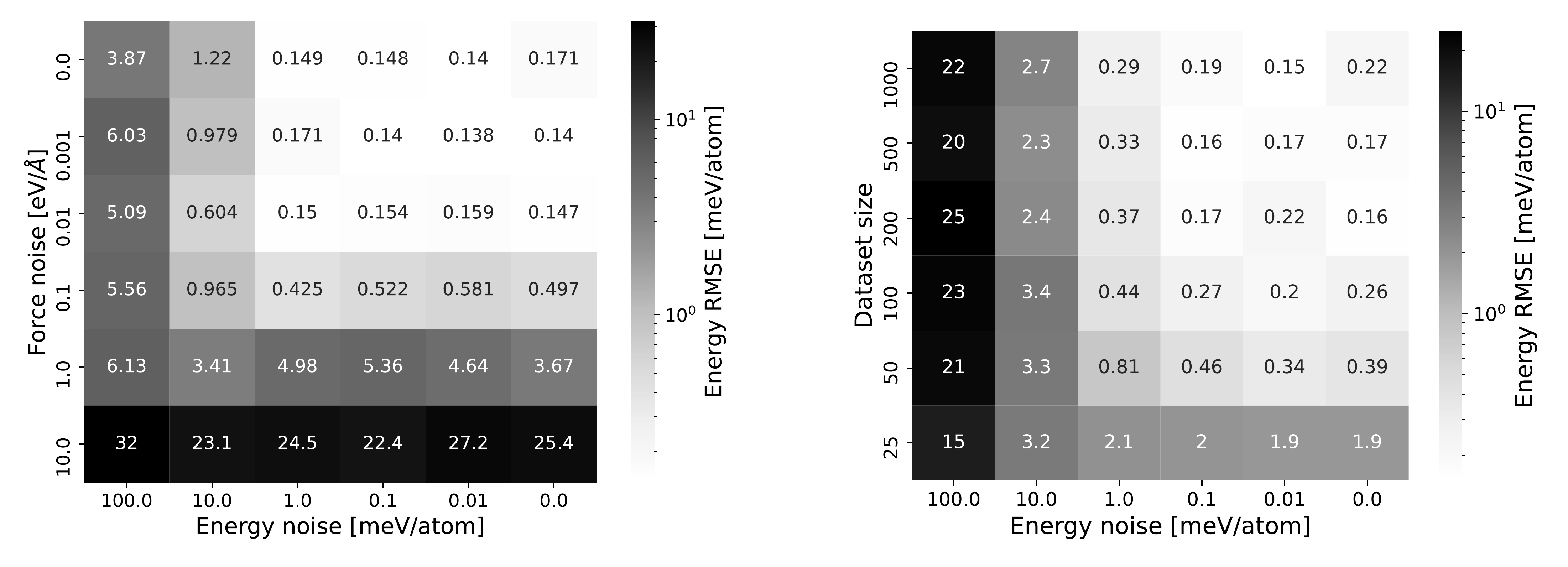}
    \caption{Single water molecule:
     Energy RMSE of the MLPs trained using noisy labels and different training dataset sizes. \textbf{Left} Energy RMSE as function of the noise level (see text) affecting the energy labels (x-axis) and the force labels (y-axis) in the training dataset. Both, the energy and force labels are used for the training. The training dataset contains 100 single water molecule structures. \textbf{Right} Energy RMSE as function of the noise level (see text) affecting the energy labels (x-axis) and the number of structures  in the training dataset (y-axis). Only the energy labels are used for the training.}
    \label{fig:1water_stat}
\end{figure*}

We first repeat the above assessment using simulated noise.
Concerning the training, we select 20 symmetry functions for hydrogen and 15 symmetry functions for oxygen with the CUR feature selection method~\cite{Imbalzano_2018} (see Appendix~\ref{app:symfunc_select}).
In this case, we also consider the possibility to train the MLP \emph{without} the use of the forces.
In this case, it is interesting to also assess the dependence of the RMSE on the dataset size.
The results of both, the training with and without using the forces, are shown in Fig.~\ref{fig:1water_stat}.
We observe qualitatively similar behavior as for the bulk water.

In view of the non-trivial computational cost of an electronic structure calculation on a quantum computer, we aim to reduce the number of configurations in the training and validation dataset as much as possible, using the CUR decomposition~\cite{Imbalzano_2018}.
In Fig.~\ref{fig:1water_stat} (right panel) we observe that we can reduce this number down to 100 configurations in the training set, without a noticeable increase in the RMSE.

In terms of the dependence on the energy noise, the behaviour of the RMSE in the training with and without using the forces is qualitatively the same, with an energy noise threshold of 10 meV/atom.
However, there is a small advantage if the forces are used in the training.
The calculation of the forces with VQE was already proposed by Sokolov et al.~\cite{Sokolov_2021}. 
However, due to technical limitations and the fact that the training without the forces is possible, we focus on the energy noise threshold in the following discussion.

The next step is to assess the number of shots $M$ (see Sect.~\ref{ss:stat_err}) required to achieve the desired accuracy.
We use the energy noise threshold of 10 meV/atom.

To estimate the number of total measurements $M$ to reach a certain accuracy $\mathcal{E}$ in the energy estimation of an $N$-qubit Hamiltonian we consider the probability $p\left(\delta < \mathcal{E} \right)$ that the deviation of the energy estimate to the ground state energy $E_0$ is smaller than the desired accuracy.
Following Ref.~\onlinecite{Torlai_2020}, this probability is given by
\begin{equation}\label{eq:qc_resource_estimation}
    p\left( \delta <  \mathcal{E} \right) = \mathrm{Erf} \left(  \mathcal{E} \sqrt{S / 2 \sigma^2 \left[ H \right] } \right) \, ,
\end{equation}
where $S = M / K$ is the number of measurements per Pauli operator $\hat{P_k}$ and $\sigma^2\left[H\right]$ is the measurement variance of the Hamiltonian.
The estimate for the total number of measurements is then given by the number of measurements that is required to reach $p\left(\delta <  \mathcal{E}\right) \approx 1$.

A loose upper bound of the resource estimation $p_{\max}$ can be obtained by determining the variance in the equation above with
\begin{equation}\label{eq:qc_sigma_max}
    \sigma^2_{\max} \left[ H \right] = \left( \sum_k \left| c_k \right| \right)^2 \,,
\end{equation}
where $c_k$ are the coefficients of the qubit Hamiltonian in Eq.~\eqref{eq:paulis_decomposition}~\cite{Wecker_2015}.
However, a more realistic estimate should be performed by directly emulating the quantum measurements process,
\begin{equation} \label{eq:qc_sigma_qc}
    \sigma^2_{\text{qc}} \left[ H \right] = \sum_{k=1}^{K} \left| c_k \right|^2 \sigma^2 [ \hat{P}_k ] \, ,
\end{equation}
where $\sigma^2 [ \hat{P}_k ]$ is the variance of the samples obtained from the measurement of the Pauli string $\hat{P}_k$.

The water molecule consists of three atoms and therefore the desired accuracy to be inserted in the previous formula is $\mathcal{E} = 30$ meV, which is comparable with chemical accuracy (1.6 mHa $\approx$ 43.5 meV).

We then define the second quantized Hamiltonian using the molecular orbitals obtained from the minimal STO-3G atomic basis set.
The fermion-to-qubit mapping is then achieved using
the parity mapping~\cite{Bravyi_2017}.
This results in a Hamiltonian encoded on 12 qubits.
We further reduce this requirement down to 9 qubits by exploiting the mapping-specific two-qubit reduction, the planar structure of the molecule (this feature holds even in the presence of distortions) as well as
the freezing of the core orbitals in the oxygen atom.
For a water molecule randomly chosen from the used training dataset, the resulting 9 qubit Hamiltonian is made of $K = 1027$ Pauli string operators.

\begin{figure}[ht]
    \includegraphics[width = 0.85 \columnwidth]{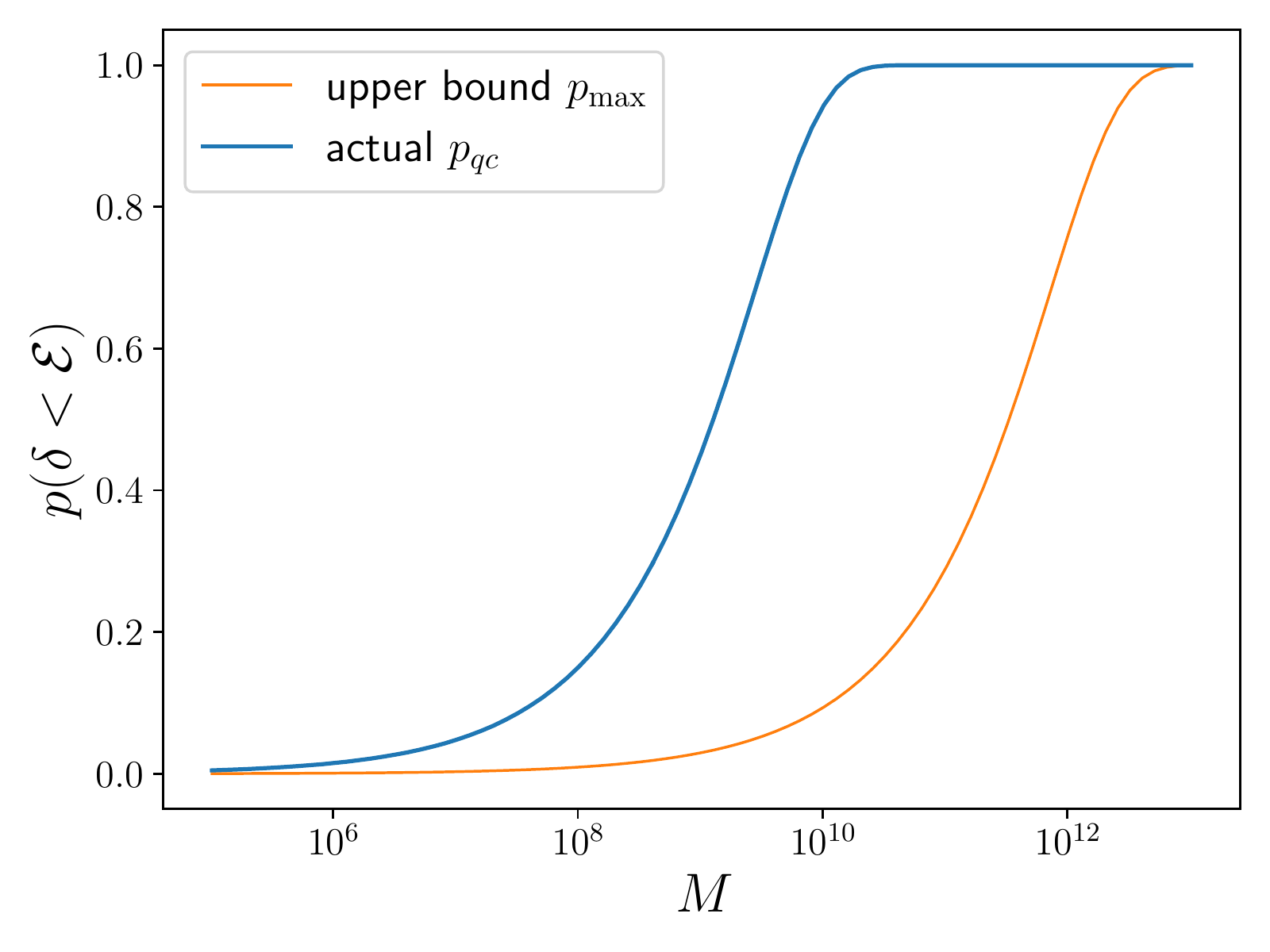}
    \caption{Probability of obtaining an energy estimate with a statistical error $\delta$ smaller than $\mathcal{E}$ as a function of the number of measurements. Here $\mathcal{E}$ is 30 meV for the single water molecule model (see text), and the probability reaches $p\approx 1$ when the number of shots is about $10^{10}$ using the standard Pauli measurement technique (blue line). The upper bound as defined in Eq.~\eqref{eq:qc_sigma_max} would exceed $10^{12}$ (orange line).}
    \label{fig:results_resource_estimation_h2o}
\end{figure}

The probability $p(\delta < \mathcal{E})$ that the deviation of the ground state energy estimate of the H$_2$O molecule to the exact ground state energy is less than $\mathcal{E} = 30$ meV is shown in Fig.~\ref{fig:results_resource_estimation_h2o}. 

From Fig.~\ref{fig:results_resource_estimation_h2o}, we observe that a probability of $p(\delta < \mathcal{E}) \approx 1$ is reached for a total number of about 10$^{10}$ measurements.
However, advances in quantum measurement protocols are expected to improve this estimate by some orders of magnitude~\cite{Kandala2017,Torlai_2020,Crawford2019,Gokhale2019,Izmaylov2019,garcia2021learning}.

\subsubsection{H$_2$-H$_2$ cluster}
This model system features two hydrogen molecules, with intramolecular distances sampled from a Gaussian distribution having mean $\mu = 1.42$ Bohr and standard deviation $\sigma = 0.03$ Bohr.
The intermolecular distances are instead sampled from a skewed distribution with two Gaussian tails of different widths.
This corresponds to distances between about 4.5 Bohr and 10 Bohr and a mean value of 6.0 Bohr.
Their respective molecular orientations are also sampled randomly.
This system is particularly challenging, since it is either unbounded or weakly bounded, depending on the level of theory\cite{doi:10.1063/1.2826340}.

We perform the same kind of assessment and investigate the RMSE on the energy and the forces as a function of the strength of the artificially generated statistical noise. 
We use a training dataset of 1000 configurations, and we compute the label using DFT, applying the PBE functional with D3 corrections.
The mean energy of the dataset is $-15.8066$ eV/atom with a standard deviation of 2.75 meV/atom.
The standard deviation in the force labels is $0.318$ eV/\AA.

\begin{figure}
    \centering
    \includegraphics[width=\columnwidth,trim=25 0 570 0,clip]{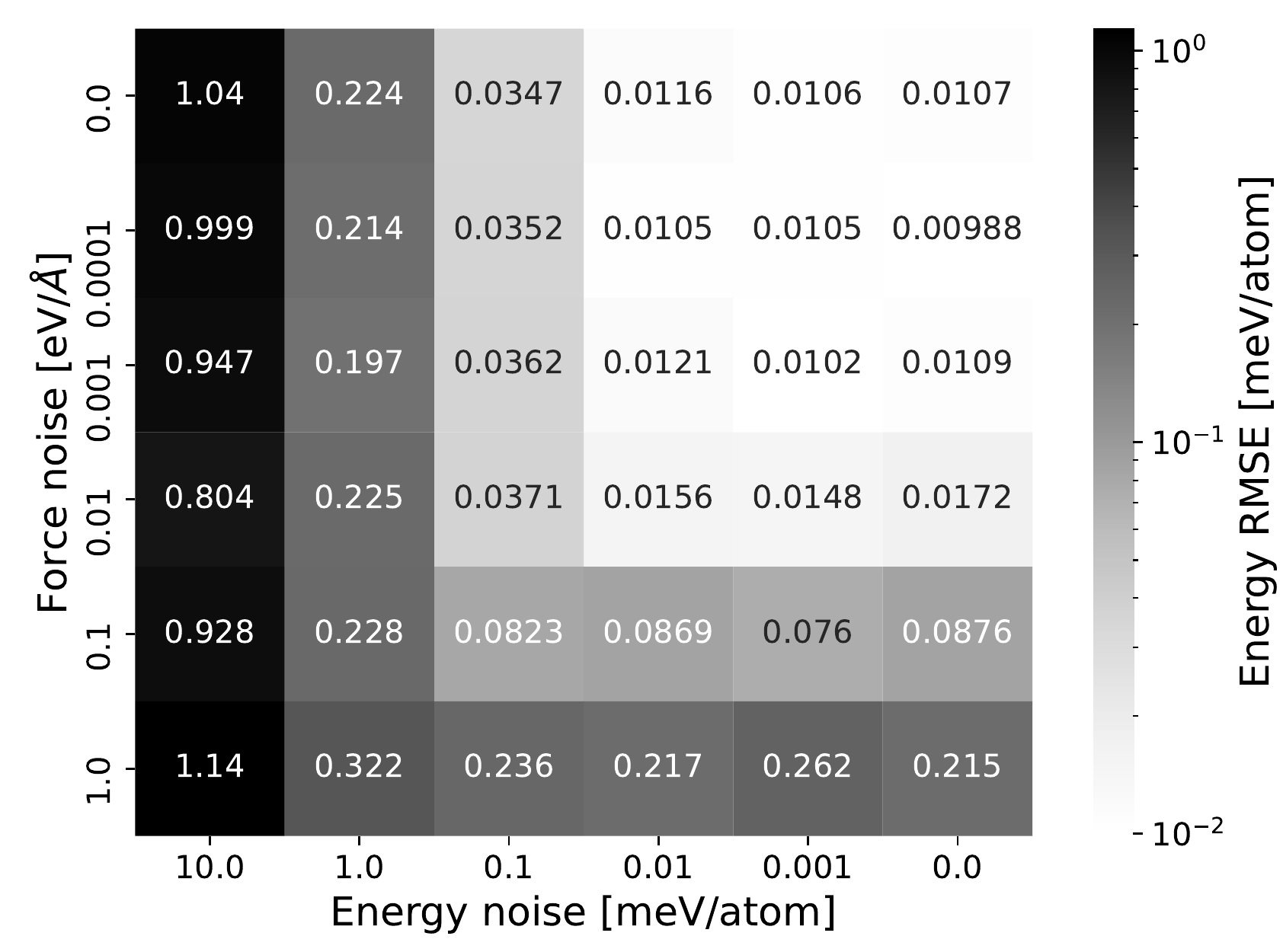}
    \caption{H$_2$-H$_2$ cluster: Energy RMSE as function of the noise level (see text) affecting the energy labels (x-axis) and the force labels (y-axis) in the training dataset.}
    \label{fig:results_resource_estimation_h2_h2}
\end{figure}

Fig.~\ref{fig:results_resource_estimation_h2_h2} shows qualitatively similar results as for the bulk water and the single water molecule cases. 
However, this time the energy noise threshold to be met in order to get an RMSE comparable to the noiseless label case is about $0.1$ meV/atom.
This target is more demanding compared to the single water molecule case as the energy scale of the bounded cluster is also much smaller (about 1.9 meV/atom).
For this reason, the number of shots required to compute the energy within the error of 0.1 meV/atom is about $10^{12}$.

\subsection{Resilience against optimization errors}
\label{ss:varnoiseresult}

In this section, we discuss the kind of error we can expect when using variational approaches for the calculation of energy expectation values.
More specifically, we consider the case when the electronic structure calculations are not fully converged. 
In particular, we test two types of variational circuits: the unitary couple cluster ansatz and the heuristic ansatz.

Clearly, the MLP is not supposed to improve upon the level of theory at which the dataset has been computed.
Therefore, here we focus on the impact of unconverged variational optimization in the training.
For instance, an ansatz featuring many variational parameters can be more difficult to optimize compared to others, thus producing a dataset with scattered labels.

We compare the performance of UCCSD\cite{Cai_2020} with a heuristic ansatz, the so-called RY-CNOT circuit, with linear connectivity and depth 24, meaning that the circuit features 24 repeating subunits of RY single qubit gates and a cascade of CNOT (or CX) gates, which represent an entangling block.
This depth was necessary to obtain results comparable with the UCCSD ansatz.
These results refer to the 9 qubit single water molecule model introduced above.
The UCCSD ansatz features 58 variational parameters with a total CNOT gate count of 4056, while the heuristic ansatz contains 225 parameters but a less complex circuit, made of 192 CNOT gates.

\begin{figure*}[ht]
    \includegraphics[width = 1 \textwidth]{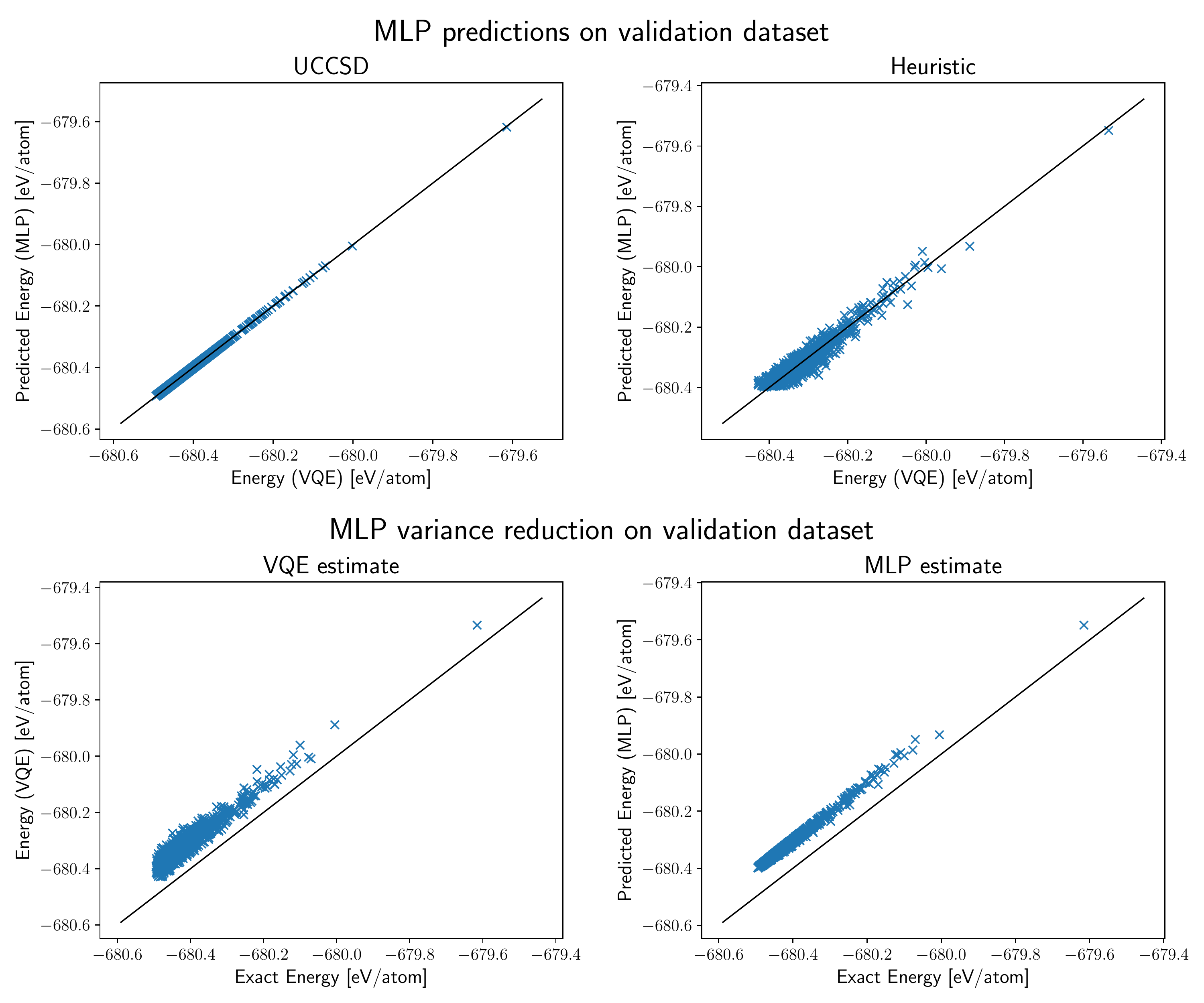}
    \caption{\textbf{Top} MLP prediction on the validation dataset for a single water molecule model, where the energy labels have been computed using VQE with the UCCSD ansatz (left) and the heuristic ansatz (right). The VQE error is present also in the validation dataset labels.
    In the bottom panels we plot both data series as a function of the exact energy instead.
    \textbf{Bottom left} VQE energy labels for the validation dataset plotted against their exact values. The positive offset shows the residual variational error of the ansatz, while the fluctuations around it are due to the optimization noise, namely the energy of some configurations is optimized better compared to others.
    \textbf{Bottom right} MLP energy predictions for the validation dataset plotted against their exact values. 
    While the MLP (correctly) cannot improve the average variational error of the ansatz, it strongly reduces the fluctuations.
    The data of the bottom panels refer to the heuristic ansatz only.}
    \label{fig:mlp_water_ansatz}
\end{figure*}

In Fig.~\ref{fig:mlp_water_ansatz} we observe that the MLP trained on datasets coming from the two different variational circuits give good results on the respective validation dataset, reaching an RMSE of 0.212 meV/atom and 21.2 meV/atom for the UCCSD and the heuristic ansatz, respectively.
As expected, the UCCSD ansatz outperforms the heuristic ansatz, since the circuit optimization proceeds smoothly in that case.

It is important to stress that the energies in the validation dataset, reported in the top panels of Fig.~\ref{fig:mlp_water_ansatz} (i.e. on the \emph{x}-axis), are also computed with VQE, meaning that they are also affected by the same optimization errors.
As it will become clear in the following, this explains, for the most part, the deviations in the top-right panel of Fig.~\ref{fig:mlp_water_ansatz}.

If we focus on the heuristic ansatz and compare the energy labels obtained by VQE with the MLP predictions on the same validation dataset, but using the exact energies as a benchmark, we observe that the MLP fit achieves a significant reduction of the energy variance (i.e, less scattered points).
Indeed, the RMSE of the validation dataset compared to the exact energy benchmark (obtained via exact diagonalization) is 96.0~$\pm$~63.9~meV/atom while the RMSE of the MLP on the same benchmark is 95.5~$\pm$~32.1~meV/atom.
Therefore, not only we were able to train successfully an MLP from noisy training and validation datasets, but the resulting MLP features a smoother energy landscape compared to the direct VQE calculations.
This property is obviously required in MD applications as an artificially corrugated potential implies unphysical, noisy, forces.

\subsection{Resilience against hardware noise}
\label{ss:hwnoiseresult}

In this section, we assess the third type of errors, which is due to the uncorrected hardware errors typical of state-of-the-art noisy quantum computers. 
Similar to the variational noise discussed in section \ref{ss:statnoiseresult}, we do not expect the MLP to improve upon the energies calculated under the effect of hardware noise. 
Therefore, the focus in this section is on the effect of hardware noise on the learned MLP.

Our assessment includes both noisy simulation and real hardware experiments.
Simulations are useful to investigate different levels of gate errors, also beyond the current values.
Real hardware experiments are important as they include all possible sources of errors, beyond the ones considered in the simulations.

\subsubsection{Noisy hardware simulations}
\label{sss:noisyhwsim}

We simulate the actual hardware noise using a custom Qiskit~\cite{Qiskit} noise model whose baseline parameters are derived from the calibration data of current IBM Quantum backends. 
The custom noisy backend consists of identical qubits and identical gates, meaning each type of gate behaves identically on all qubits ($\mathbb{I}$, RZ, SX, X) and all pairs of qubits (CX). 
The parameters of the custom backend are listed in Tab.~\ref{tab:custom_backend_parameters}.

We specifically focus on two types of hardware noise: gate error and readout error. 
To make the full analysis suited for hardware calculations, we will limit our investigation to a simpler model, namely the hydrogen molecule, $H_2$.

\begin{table}
    \centering
    \begin{tabular}{ l | c }
        Parameter & Baseline value \\
        \hline
        Thermal relaxation time T1 & 100 $\mu$s \\
        Dephasing time T2          & 100 $\mu$s \\
        Qubit frequency            & 4.77 GHz   \\
        Qubit anharmonicity        & -0.334 GHz \\
        Readout error ($| 0 \rangle$ instead of $| 1 \rangle$) & 4 \% \\
        Readout error ($| 1 \rangle$ instead of $| 0 \rangle$) & 2 \% \\
        1-qubit gate error*        & 0.03 \% \\
        1-qubit gate time*         & 35.6 ns \\
        CX gate error              & 1 \% \\
        CX gate time               & 430 ns \\
    \end{tabular}
    \caption{Baseline parameters for custom noise backend. 
    The values are either taken directly from a specific qubit (frequency and anharmonicity) or inspired by an average of different IBM Quantum devices (all remaining values). \\
    \footnotesize{*the RZ gate is applied virtually and therefore the gate error and time are both 0.}}
    \label{tab:custom_backend_parameters}
\end{table}

\paragraph{Gate error.}
\label{par:gate_err}
We model the gate error using coherence limited fidelity for individual gates, i.e. assuming that the reported gate error is due solely to thermal relaxation and dephasing effects, parameterized with the thermal relaxation time T1 and the decoherence time T2, respectively. 
While this simplified scenario does not entirely reflect the actual experimental conditions -- where other effects (e.g., coherent control and calibration errors) and noise channels (including correlated multi-qubit noise) may be present -- it is nevertheless sufficient to capture the dominant behavior of current noisy processors without complicating the analysis.
The baseline value for both T1 and T2 is 100~$\mu$s (see Tab.~\ref{tab:custom_backend_parameters}). 
However, in simulations it is also possible to assess different scenarios that would correspond to future expected technical improvements in device fabrication. 
To this end, we systematically increase T1 and T2 to investigate the effect of a future gate error reduction on the calculation of the energies. 
More specifically, we extend T1 to a maximum value of 2~ms, which is a realistic prediction for the next years according to recent hardware developments.

\begin{figure}
    \centering
    \includegraphics[width=\columnwidth]{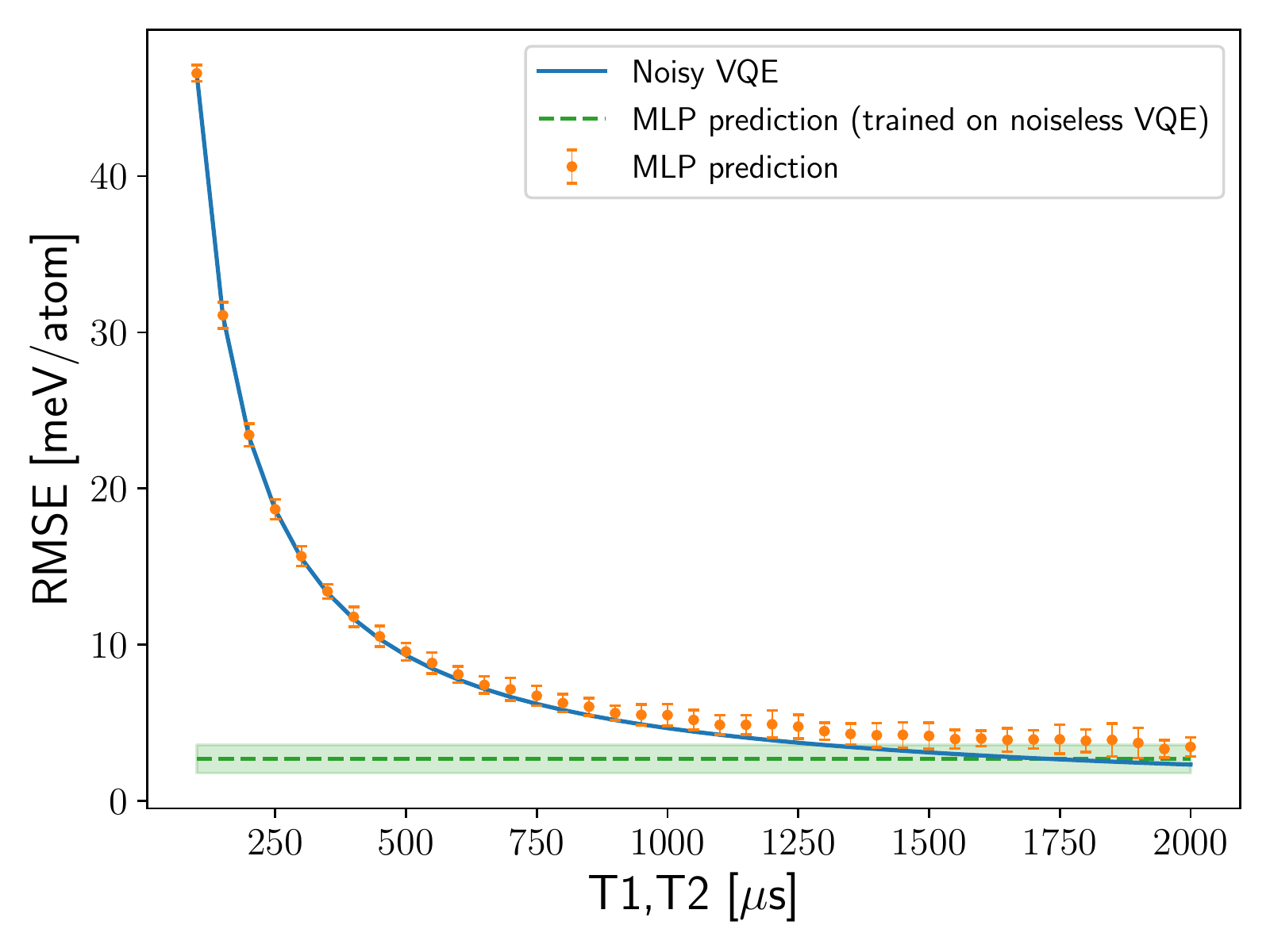}
    \caption{Average energy RMSE at different levels of gate errors. 
    The gate errors are characterized by the thermal relaxation time T1 and the dephasing time T2, which are set to the same value and varied simultaneously. 
    The blue solid line shows the energy RMSE of the validation configurations, where the energies are obtained at the corresponding level of gate errors and compared to the respective noiseless energies. 
    Each orange point is an average of 20 MLPs that were trained on different training datasets. The error bars shows one standard deviation of the energy predictions.
    The green dashed line serves as a reference, and shows the average energy RMSE of the MLP predictions where no gate errors were present in the energy calculations of the reference datasets.}
    \label{fig:hw_gate_error}
\end{figure}

For this analysis, we randomly create 20 training datasets and one validation dataset for the hydrogen molecule, each with 20 configurations characterized by different bond lengths. 
Each molecular configuration is randomly rotated in space and features an intramolecular bond distance in the range [0.6, 4.2] Bohr.
The H$_2$ wavefunction is encoded in the STO-3G basis set using the parity mapping and the mapping specific two-qubit reduction, which results in a Hamiltonian on 2 qubits. 
The VQE calculations feature a simple variational ansatz tailored to the system, that contains only one variational parameter and one CNOT gate.
For each dataset, we calculate the energy of the configurations at different levels of the gate errors and train an MLP on each set of labels. 
As a reference, we also train an MLP on each training dataset with noiseless energies. 
In Fig.~\ref{fig:hw_gate_error}, we report the energy RMSE at different gate errors for the configurations in the validation dataset (solid blue line), the average energy RMSE of the MLP predictions (orange dots) and the average energy RMSE of the reference (noiseless) MLP predictions (dashed green line). 
All RMSE values are given with respect to energies obtained with noiseless VQE calculations.

We first notice that the MLP predictions closely follow the blue line (for noisy VQE), and we therefore conclude that the MLPs faithfully learn the \emph{noisy} potential energy landscape.
As expected, reducing the gate error leads to more accurate MLP in the absolute sense.
The crossing point, at which the error due to the gate noise gets smaller than the model error of the MLP, is at about T1 $\approx$ 1.75 ms.
Above this value, the MLP error saturates to the model error. 

Some comments are in order: on a positive take, these results show that, in principle, even a finite \emph{gate noise} can produce MLPs which are as accurate as the best MLP trained on noiseless data.
On the other side,
the hydrogen molecule is one of the simplest systems we can study, and the circuit used is very shallow.
Larger molecules will require much deeper circuits, and therefore, we expect that the effect of the gate errors will increase significantly.
Therefore, this assessment represents a \emph{best-case scenario} concerning MLP training on quantum data in the non fault-tolerant setting.

\paragraph{Readout error.}
\label{par:readout_err}
The second type of error we simulate is the readout error.
The baseline parameters for the readout errors are listed in table~\ref{tab:custom_backend_parameters}.
The readout error is best probed by emulating the measurement process.
To highlight readout inaccuracies, we suppress the statistical fluctuations (see Sec.~\ref{ss:stat_err}) using 10$^5$ shots per circuit to measure the energy.

\begin{figure}
    \centering
    \includegraphics[width=\columnwidth]{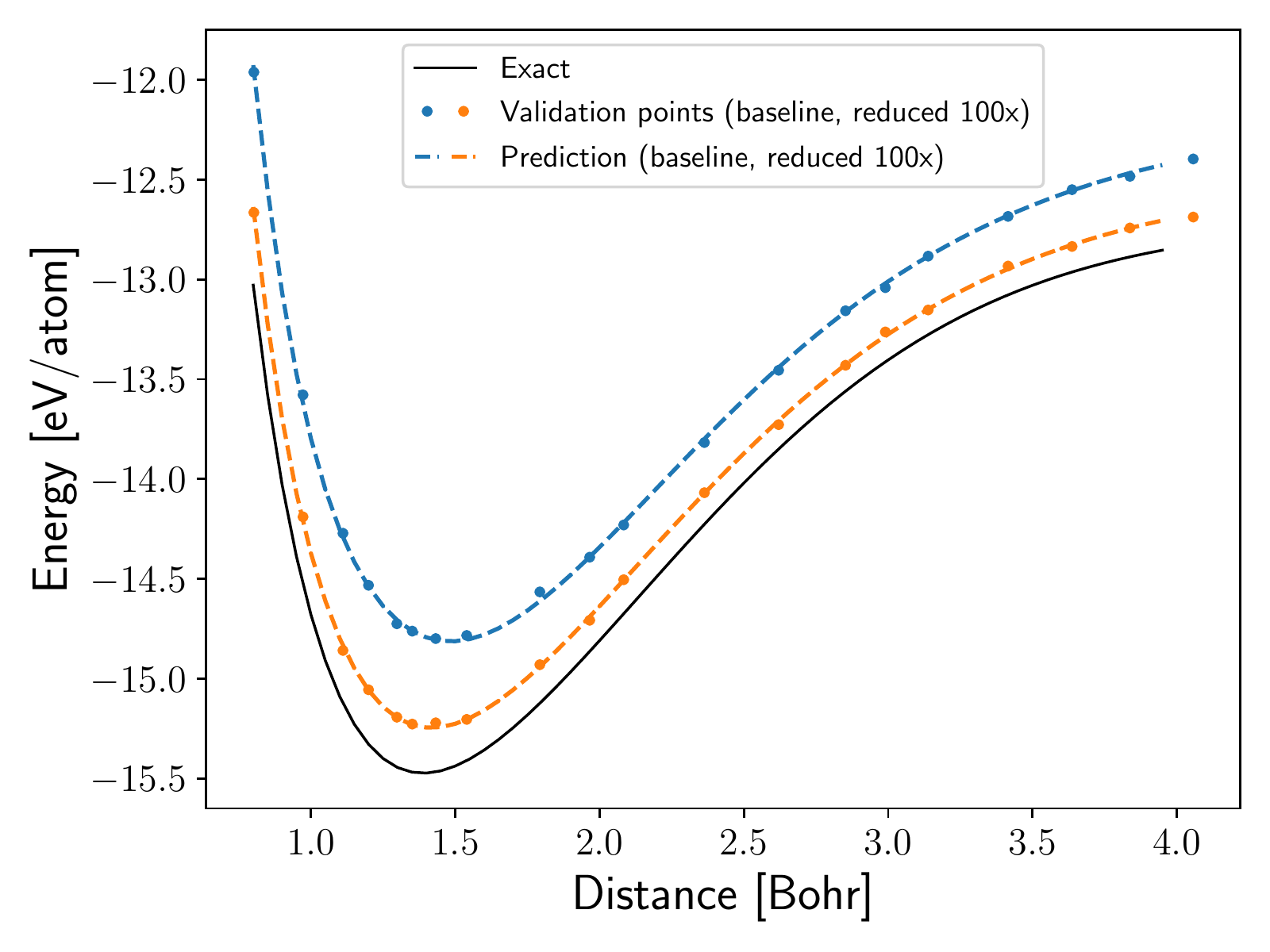}
    \caption{Predicted hydrogen molecule dissociation path at different levels of readout error assumed in the calculation of the reference energies. 
    The dashed lines show the predictions by the MLPs, the dots show the energies of the configurations in the validation datasets and the black line shows the dissociation path obtained by exact diagonalization. 
    The baseline readout error values used for the data in blue are listed in Tab.~\ref{tab:custom_backend_parameters}. 
    For the data in orange, the readout error is reduced by a factor of 100.}
    \label{fig:hw_readout_error}
\end{figure}

In this section, we simulate the readout error at baseline level and at a level where the error is reduced by a factor of 100 to create a training and validation dataset of the hydrogen molecule, each with 20 configurations. 
For demonstration purposes, we also train and evaluate an MLP on the resulting datasets. 
Their performance is reported in Fig.~\ref{fig:hw_readout_error}. 
In the figure, the black solid line shows the exact energies of the hydrogen molecule dissociation path, the dashed lines show the predictions of the trained MLPs, and the dots show the energies of the configurations in the validation datasets. 
The MLPs achieve an energy RMSE of 19.7~meV/atom and 13.5~meV/atom on the validation dataset, having the baseline and the reduced readout error, respectively. 
Compared to the exact energy, they show an error of 607~meV/atom and 210~meV/atom.
As expected, reducing the readout error leads to more accurate energy estimations and therefore to more accurate MLPs.

\subsubsection{Hardware experiments}
\label{sss:hw_exp}

Finally, we also run experiments on IBM Quantum superconducting processors, where all actual error sources are present.
We run the hardware calculations for a training and validation dataset of the hydrogen molecule, each with 20 configurations.
For each configuration we run 4, 5 and 10 VQE runs on the IBM Quantum devices \textit{ibmq\_toronto}, \textit{ibmq\_bogota} and \textit{ibmq\_manila}, respectively. 
All these quantum processors feature a  quantum volume of 32.
The final energy label is obtained by averaging over the different experiment realizations, after excluding clear unconverged runs.
Data measured on different devices contribute to separate datasets. 
All energy expectation values are computed with 8192 measurements (or shots).

\begin{figure*}
    \centering
    \includegraphics[width=\textwidth]{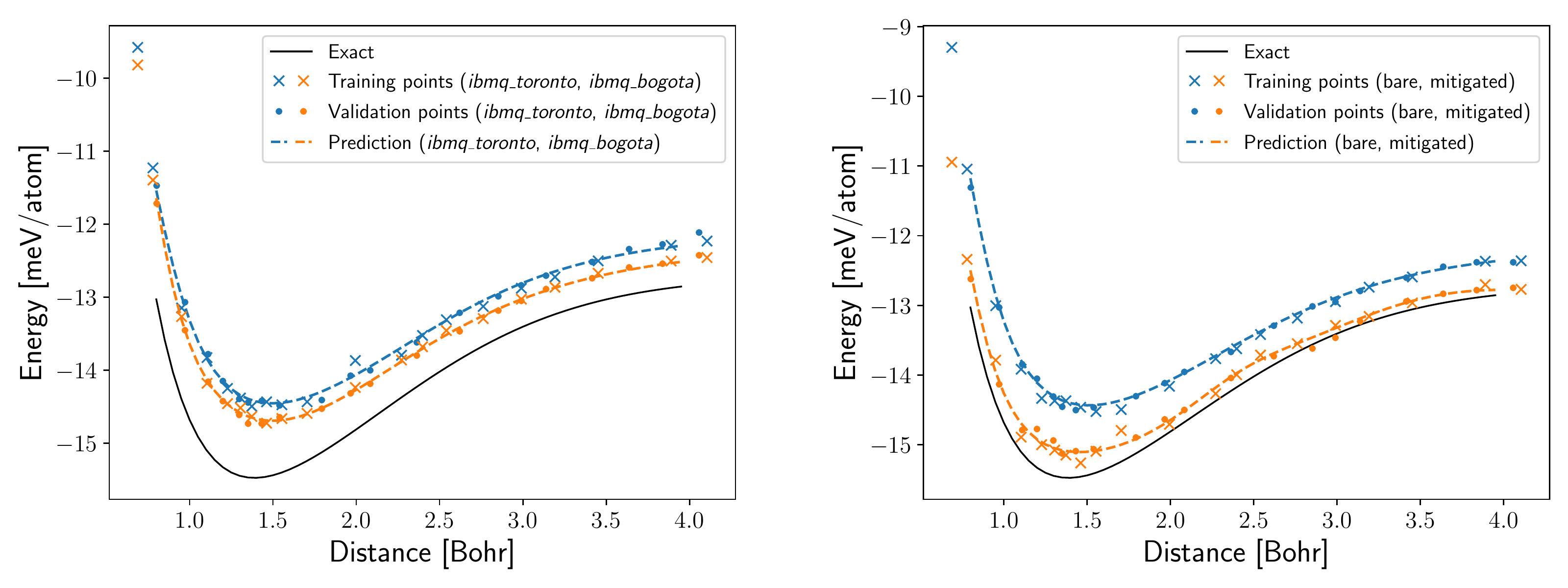}
    \caption{\textbf{Left} Prediction of hydrogen molecule dissociation path by an MLP that was trained and evaluated on datasets obtained with the IBM Quantum devices \textit{ibmq\_toronto} (blue) and \textit{ibmq\_bogota} (orange). The energies in the training and validation dataset are a filtered average over 4 (5) VQE runs for \textit{ibmq\_toronto} (\textit{ibmq\_bogota}).
    \textbf{Right} Prediction of hydrogen molecule dissociation path by an MLP that was trained and evaluated on datasets obtain with the IBM Quantum device \textit{ibmq\_manila} using no readout error mitigation (blue) and using readout error mitigation (orange). The energies in the training and validation dataset are a filtered average over 10 VQE runs.}
    \label{fig:hw_exp}
\end{figure*}

The results are summarized in Fig.~\ref{fig:hw_exp}. 
The plot on the left reports the curves obtained with \textit{ibmq\_toronto} (blue) and \textit{ibmq\_bogota} (orange), while the plot on the right shows the results for \textit{ibmq\_manila} (blue).
In all cases, the predictions of the MLPs closely follow the data points of the training and validation dataset (cross and point markers, respectively). 
The difference between the curves is due to the different properties of each device, with \textit{ibmq\_bogota} outperforming \textit{ibmq\_toronto}. 

With \textit{ibmq\_manila}, we also apply measurement error mitigation techniques; to this end,  
we apply the full calibration matrix method~\cite{Bravyi_2021} on 10 additional VQE runs of each selected configuration, where the calibration matrix is refreshed every 30 minutes.
To estimate the final energy of each configuration, we apply the same procedure as described above; as usual, we neglect unconverged VQE runs and average over the remaining energy measurements.
The energy estimates using the MLP trained on the obtained datasets are shown in Fig.~\ref{fig:hw_exp} in the panel on the right.
We observe that the mitigated energies are much closer to the exact energies, in agreement with the error simulation of section~\ref{par:readout_err}.

We stress once more that the goal of training an MLP from quantum data is not hardware error mitigation \textit{per se}, but rather to obtain a smooth and re-usable interpolation of the noisy data.
The computational gain compared to the straightforward molecular dynamics approach of Ref.~\onlinecite{Sokolov_2021} even when performed on the same single molecule, is straightforward. 
The standard method requires a new VQE calculation for each iteration, while the training of an MLP in this specific case only requires order $\mathcal{O}(10)$ single point VQE runs.
Moreover, in Ref.~\onlinecite{Sokolov_2021} a costly Lanczos error mitigation scheme was sometimes needed, making every single-point VQE calculation $\mathcal{O}(10^2)$ times more expensive compared to the present work, where the stability of the dynamics is ensured by the smoothness of the MLP surface.
Finally, this cost needs to be multiplied by the total number of time steps of the MD. To summarize, for a 100 fs simulation of the $H_2$ molecule, assuming a time step of 0.2 fs, the total cost of a stable quantum-powered MLP simulation is now reduced by a factor of $10^5$ compared to the straightforward approach.

\section{Conclusions}
\label{s:conclusion}

We propose the usage of classical machine learning potentials (MLP) trained on quantum electronic structure data to enable large-scale materials simulations.
The motivation behind this is quite simple: while quantum computing algorithms can outperform their classical counterpart for electronic structure problems, they still feature a polynomial runtime (possibly with a large prefactor) that can still prevent applications to bulk materials.

MLPs have been successfully introduced in materials simulations powered by classical approximate electronic structure solvers, enabling truly large-scale and equilibrated simulations\cite{Deringer_2021,Cheng_2020}.
Here, we assess the trainability of an MLP using quantum data obtained from a variational calculation.
In particular, we study the impact of three types of noise that are characteristic to the quantum algorithm: the statistical noise, the optimization error, and the hardware errors.

These errors impact the training and validation energy labels as
\begin{equation}
E_{\text{label}} = E_{\text{true}} + \Delta + \eta
\end{equation}
where $\Delta$ is a systematic error and $\eta$ is a fluctuation around this offset.
While the MLP is not intended to compensate for any systematic error $\Delta$, it can greatly mitigate the random fluctuations affecting the labels.
These  may arise from the statistical error in the evaluation of the energy and forces estimator, the VQE optimization error that may affect some dataset points more than others, as well as any non-systematic component of the hardware noise.

Here we use an MLP based on state-of-the-art neural-network potentials, and show that there exist a threshold value for the noise strength, such that we can achieve a training as good as in the noiseless case.
The resulting MLP features a smooth energy surface that would allow for stable molecular dynamics simulations or structural optimizations.

We substantiate our research through the simulations of the separate sources of noises.
We finally generate training and validation datasets using actual quantum hardware and obtain the first MLP trained with electronic structure calculations using a real quantum computer.

While in our assessment we consider a neural-network type of MLP, next research directions include the possibility of using kernel-based models, which tend to have a better performance when only few training data is available~\cite{doi:10.1021/acs.chemrev.0c01111}.

\section{Acknowledgements}
I.T. acknowledges the financial support from the Swiss National Science Foundation (SNF) through the grant No. 200021-179312. IBM, the IBM logo, and ibm.com are trademarks of International Business Machines Corp., registered in many jurisdictions worldwide. Other product and service names might be trademarks of IBM or
other companies. The current list of IBM trademarks is available at \url{https://www.ibm.com/legal/copytrade}.

\newpage
\appendix

\section{Dataset normalization}
\label{app:dataset_normalization}

Normalizing the labels in the training dataset is a common practice to improve the training of a Neural Network~\cite{Singraber_2019_training}.
In the case of a machine learning potential (MLP) the normalization defines internal units which are independent of a physical unit system.
Fortunately the normalization process is integrated into the training procedure of \texttt{n2p2}~\cite{n2p2}.
Given a dataset of atomic structures with energy (E) and force (F) the normalization transformation is parameterized by three parameters, the mean energy per atom $\langle E \rangle$, a conversion factor for the energies $c_{\text{energy}} = 1 / \sigma_E$ and a conversion factor for distances $c_{\text{length}} = \sigma_F / \sigma_E$, where $\sigma_E$ and $\sigma_F$ are the standard deviation of the energy and the forces, respectively.
Applying the transformation
\begin{align}
    E^*_i = \frac{E^i_{\text{Ref}} - N^i_{\text{a}} \langle e \rangle}{\sigma_e} \,, \\
    F^* = \frac{c_{\text{energy}}}{c_{\text{length}}} F_{\text{Ref}} \,,
\end{align}
to each configuration in the dataset, ensures that the transformed labels ($E^*, F^*$) have zero mean and unit standard deviation~\cite{Singraber_2019_training}, i.e. $\langle E^* \rangle = 0$, $\sigma_E^* = 1$ and $\sigma_F^*$ = 1 (the forces should already have zero mean).

The normalization is successful as long as we input both, the energies \emph{and} the forces. 
However, using the VQE algorithm we only calculate the energy of the atomic configurations and set all the forces to zero.
Concerning the training this is no issue, as the training on only the energy labels is supported by \texttt{n2p2}.
However, during the normalization process, the conversion factor for distances is set to $c_{\text{length}} = 0$ ($\sigma_F = 0$).
This leads to problems in subsequent steps of the training process.
Therefore, if the forces are not available for the normalization process, we manually set $c_{\text{length}} = 1$.

\section{Symmetry functions}
\label{app:symfunc}

Two types of symmetry functions are used in this manuscript.
The first is the radial symmetry functions class.
For an atom labelled with index $i$ the radial symmetry function is defined as
\begin{equation}\label{eq:nnp_radial_sf}
    G_2^i = \sum\limits_j e^{- \eta \left( R_{ij} - R_s \right)^2} \cdot f_c(R_{ij}) \,, 
\end{equation}
where $\eta$ determines the width and $R_s$ the position of the Gaussian.

The cutoff function $f_c(R_{ij})$ ensures that the value and the derivative of the symmetry function go to zero if the distance $R_{ij}$ between the central atom and a neighboring atom is bigger than the cutoff radius $R_c$. 

The cutoff function that we used for the MLPs trained in this paper is
\begin{equation}\label{eq:cutoff_function}
    f_c(R_{ij}) = \begin{cases}
                  \tanh^3 \left[ 1 - R_{ij}/R_c \right] & \text{for $R_{ij} \leq R_c$} \\
                  0                                     & \text{for $R_{ij} > R_c$}
                  \end{cases} \,.
\end{equation}
A list of other cutoff functions can be found in Ref.~\onlinecite{Singraber_2019}.

The second type of symmetry functions we consider are angular symmetry functions which are a sum of three-body terms.
It is defined as
\begin{align}\label{eq:nnp_angular_sf}
    G_3^i & = 2^{1 - \zeta} \sum\limits_j \sum\limits_{k \neq j} \left( 1 + \lambda \cdot \cos \theta_{ijk} \right)^\zeta \nonumber \\
          & \times e^{-\eta \left(R_{ij}^2 + R_{ik}^2 + R_{jk}^2 \right)} \nonumber \\
          & \times f_c(R_{ij}) f_c(R_{ik}) f_c(R_{jk}) \,,
\end{align}
where $\theta_{ijk} = \arccos \left(\frac{\mathbf{R_{ij}} \cdot \mathbf{R_{ik}}} {R_{ij} \cdot R_{ik}} \right)$ is the angle between the three atoms $i$, $j$ and $k$. 
The parameters determining the shape of the function are $\eta$, $\lambda$ and $\zeta$. 
The parameter $\eta$, again, determines the width of the Gaussian part of the function. 
The parameter $\lambda$ can only take the values $1$ and $-1$ which shifts the maximum of the cosine part either to $\theta_{ijk} = 0^{\circ}$ or $\theta_{ijk} = 180^{\circ}$, respectively. The parameter $\zeta$ determines the angular resolution.

{\bf Normalization}
Similarly to the labels also the input to the NN is normalized. 
This balances the impact of the different symmetry functions on the first hidden layer in the NN.

The normalization transformation is
\begin{equation}
    G^{\text{scaled}}_i = \frac{G_i - \langle G_i \rangle}{G_{i,\max} - G_{i,\min}} \,,
\end{equation}
which centers the symmetry functions $G_i$ with their mean $\langle G_i \rangle$ and rescales them to the interval $[-1, 1]$~\cite{Singraber_2019_training}.

{\bf Forces}
The force component $F_{i,k}$ of atom $i$ is calculated from the total energy by taking the derivative with respect to the component $k$ of the position $\vec{R}_{i}$ of the atom,
\begin{equation}\label{eq:nnp_forces_definition}
    F_{i,k} = - \frac{\partial E_{\text{NNP}}}{\partial R_{i,k}} = - \sum\limits_{i=1}^{N_{\text{atom}}} \frac{\partial E_i}{\partial R_{i,k}} \,.
\end{equation}
This expression can be evaluated by applying the chain rule,
\begin{equation}\label{eq:nnp_force}
    F_{i,k} = - \sum\limits_{i=1}^{N_{\text{atom}}} \sum\limits_{j=1}^{N_{\text{sym},i}} \frac{\partial E_i}{\partial G_{ij}} \frac{\partial G_{ij}}{\partial R_{i,k}} \,,
\end{equation}
where $N_{\text{sym},i}$ is the number of symmetry functions for atom $i$ and $G_{ij}$ the $j$-th symmetry function of atom $i$.
The first partial derivative is given by the functional form of the NN.
The second partial derivative is given by the functional form of the symmetry functions and can be calculated analytically. 
For the two symmetry function types given above, the derivatives can be found in the Supporting Information of Ref.~\onlinecite{Singraber_2019}.

\section{Automatic selection of symmetry functions}
\label{app:symfunc_select}

This section provides a short review of the method to automatically select symmetry functions proposed in Ref.~\onlinecite{Imbalzano_2018} and adopted in this work.
The algorithm is based on a feature selection method, called the CUR decomposition, which creates a low-rank approximation of the initial feature matrix $\mathbf{X}$ in terms of its columns and rows.
The first step of the procedure is the construction of a pool of $N$ candidate symmetry functions $\{ \Phi_j \}$. 
Given a dataset of $M$ configurations $\{ A_i \}$, the feature matrix is defined as $X_{ij} = \Phi_j(A_i)$.
In the second step we then apply the feature selection to the columns (rows) of the feature matrix to select a small subset of $N'$ symmetry functions ($M'$ configurations) which capture the important structural information of the considered system.
Below the two steps are reviewed in more detail.

\medskip

The creation of candidate symmetry functions is done by generating values for the parameters that determine the shape of the symmetry functions. 
For the radial symmetry functions $G_2$ (Eq.~\eqref{eq:nnp_radial_sf}) two parameter sets are created. 
In the first set, the Gaussians are centered at the reference atom ($R_s = 0$) and have widths chosen according to 
\begin{equation}\label{eq:methods_nnp_sf_width}
    \eta_m = \left( \frac{n^{m/n}}{r_c} \right)^2 \,,
\end{equation}
where $n$ is the number of desired parameters in this parameters set and $m = 0, 1, \dots n$. 
The second set of parameters is created in the following way
\begin{align}\label{eq:methods_nnp_sf_radial_shift}
    R_{s,m} & = \frac{r_c}{n^{m/n}} \,, \\
    \eta_{s,m} & = \frac{1}{\left( R_{s,n-m} - R_{s,n-m-1} \right)^2} \,,
\end{align}
which creates a set of Gaussians that are narrow close to the reference atom and wider as the distance increases.

For the angular symmetry functions $G_3$ (Eq.~\eqref{eq:nnp_angular_sf}) only one set of parameters is created. 
The values for $\eta$ are chosen according to Eq.~\eqref{eq:methods_nnp_sf_width}, $\lambda$ takes the values $\{-1, 1\}$ and for $\zeta$ a few values on a logarithmic scale are chosen, e.g. $\{1, 4, 16\}$. 

\medskip

The method to select the most important features of a feature matrix $\mathbf{X}$ has the following form
\begin{equation}\label{eq:cur_decomposition}
    \mathbf{X} \approx \mathbf{\widetilde{X}} = \mathbf{C} \, \mathbf{U} \, \mathbf{R} \,,
\end{equation}
where $\mathbf{C}$ and $\mathbf{R}$ are matrices that consist of a subset of columns and rows of the original feature matrix $\mathbf{X}$. 
We execute the following steps for the selection of the subset of columns ($\mathbf{C}$).
\begin{itemize}[noitemsep]
    \item Calculate the singular value decomposition (SVD) of $\mathbf{X}$.
    \item Calculate an importance score for each column $c$,
        \begin{equation}\label{eq:cur_column_score}
            \pi_c = \sum\limits_{j=1}^k \left(v_c^{(j)} \right)^2 \,,
        \end{equation}
        where $v_c^{(j)}$ is the $c$-th coordinate of the $j$-th right singular vector and $k$ is the number of singular vectors that are considered for the score. 
        A value of $k=1$ is proposed for an efficient selection.
    \item Pick column $l$ with the highest importance score.
    \item Orthogonalize the remaining columns in the feature matrix $\mathbf{X}$ with respect to the $l$-th column $X_l$
        \begin{equation}\label{eq:cur_orthogonalize}
            X_j \leftarrow X_j - X_l \, (X_l \cdot X_j) / \left| X_l \right|^2 \,.
        \end{equation}
    \item Repeat the steps above on the orthogonalized matrix until the desired number of columns is reached or the error of the approximation (Eq.~\eqref{eq:cur_error}) is below a desired threshold.
\end{itemize}
The extracted columns form the matrix $\mathbf{C}$. 
Similarly, the matrix $\mathbf{R}$ can be constructed using the algorithm above to select a subset of columns from $\mathbf{X}^T$ (rows from $\mathbf{X}$). 
The matrix $\mathbf{U}$ then is defined as $\mathbf{U} = \mathbf{C}^+ \, \mathbf{X} \, \mathbf{R}^+$ ($^+$ indicates the pseudoinverse). 
The accuracy of the approximation is
\begin{equation}\label{eq:cur_error}
    \epsilon = \left\| \mathbf{X} - \mathbf{C} \, \mathbf{U} \, \mathbf{R} \right\|_F / \left\| \mathbf{X} \right\|_F \,.
\end{equation}

\section{Creating reference datasets}
For the bulk water we use a dataset that is already established in the literature~\cite{Morawietz_2016, Morawietz_2019, Singraber_2019_training}.
The dataset for the single water molecule is derived from the bulk water dataset by extracting H$_2$O atom groups. 
The initial training dataset was then created by randomly selecting 1000 configurations from the extracted H$_2$O configurations.
We could further reduce the number of configurations by applying the CUR decomposition~\cite{Imbalzano_2018} (see Sect.~\ref{app:symfunc_select}) to select for a subset of the configurations.
We found that 100 is a convenient dataset size.
The validation dataset is created in the same way, by choosing a different (distinct) set of extracted H$_2$O configurations.

\medskip

We created our own reference datasets for the H$_2$-H$_2$ cluster and the hydrogen molecule, H$_2$. 
For the creation of a reference dataset it is recommended to use the procedure reviewed in Ref.~\onlinecite{Behler_2015}.
However, the considered systems are very simple, and we expect a sufficient coverage of the configuration space already from a random sampling.
In both cases, the H$_2$-H$_2$ cluster and the hydrogen molecule, we created an initial dataset with 1000 configurations.
We also tried to reduce the number of configurations in the dataset with the CUR decomposition.
For the H$_2$-H$_2$ cluster the training accuracy got gradually worse when reducing the number of configurations, so we kept all the 1000 configurations in the training dataset.
In the case of the hydrogen molecule we found that 20 is a convenient dataset size.
For both systems the validation datasets are also created by randomly sampling from the configuration space of the respective system.

%

\end{document}